\documentclass[pdflatex,sn-nature]{sn-jnl}
\usepackage{graphicx}
\usepackage{amsmath,amssymb,amsfonts}
\usepackage{bm,braket}
\usepackage{booktabs,tabularx}
\usepackage{multirow,makecell}
\usepackage{float,placeins}
\usepackage{textcomp}
\usepackage{xcolor}

\begin{document}

\title{Logical error prediction and geometry selection for rectangular surface codes under biased circuit-level noise}

\author[1]{\fnm{Youshin} \sur{Chung}}
\author[1]{\fnm{Yujin} \sur{Kang}}
\author*[1]{\fnm{Jun} \sur{Heo}}\email{junheo@korea.ac.kr}
\affil[1]{\orgdiv{School of Electrical Engineering}, \orgname{Korea University}, \orgaddress{\street{145 Anam-ro, Seongbuk-gu}, \city{Seoul}, \postcode{02841}, \country{Republic of Korea}}}

\abstract{Realistic quantum hardware often exhibits biased noise, but at the circuit level this bias may affect only a subset of operations and need not be transferred uniformly to logical errors. The logical response to a given physical bias is therefore central to selecting resource-efficient surface code geometries. Here, we study unrotated rectangular surface codes under standard depolarizing and superconducting-inspired circuit-level noise models. We first construct geometry-dependent logical error rate models from unbiased circuit-level simulations. We then use logical sensitivities to compare the responses to localized idle-bias and global physical error perturbations, defining logical-error-type-dependent response coefficients from an unbiased reference circuit without fitting to finite-bias data. These coefficients are incorporated through effective error parameters, yielding a median fold error of 1.09 over the evaluated finite-bias conditions. Geometries selected by the model are validated through direct biased circuit-level simulations. At a physical error rate of 0.004 and a target logical error rate of $10^{-3}$, the validated rectangular codes reduce physical qubit overhead by up to 42.4\% and 26.2\% under the two noise models, respectively, relative to unrotated square codes. These results provide a practical link between circuit-level bias response and resource-efficient surface code design.}


\maketitle

\section{Introduction}\label{sec:introduction}

The surface code is a leading candidate for fault-tolerant quantum computing because it relies on local interactions in two dimensions and exhibits a high error threshold \cite{kitaev2003fault,bravyi1998quantum,raussendorf2007fault,fowler2009high}. In many physical qubit platforms, however, noise is asymmetric, with dephasing-related errors occurring more frequently than bit-flip errors \cite{iolius2022performance,seis2023improving}. Rectangular surface codes can exploit such asymmetry by independently adjusting the logical $X$ and $Z$ distances, thereby reducing unnecessary protection along the less error-prone direction \cite{lee2021rectangular,PhysRevX.13.031007}. Realizing this resource advantage requires quantitative prediction of the logical $Z$ and $X$ error rates under circuit-level noise, since the appropriate geometry is determined by their relative suppression.

Previous studies have addressed this problem under different noise assumptions. Code-capacity analyses assume perfect syndrome extraction and neglect faults introduced by stabilizer-measurement circuits \cite{watson2014logical}. Under biased code-capacity noise, Lee et al.\ showed that optimizing the two lattice dimensions can substantially reduce physical qubit overhead \cite{lee2021rectangular}. At the circuit level, Higgott et al.\ demonstrated that bias-tailored geometries can remain advantageous when syndrome-extraction faults are included \cite{PhysRevX.13.031007}. However, the physical bias of the underlying qubits is not necessarily preserved uniformly throughout a realistic syndrome-extraction circuit. In particular, bias-preserving CNOT operations are not generally available for two-level qubits \cite{guillaud2019repetition,puri2020bias,q7w6-nljp}, motivating hybrid biased--depolarizing (HBD) descriptions that distinguish physical processes retaining the underlying bias from circuit operations with effectively depolarizing faults \cite{q7w6-nljp}. Bias degradation during faulty syndrome extraction can therefore reduce the resource advantages predicted under simpler noise assumptions \cite{benito2026optimizing}.

These circuit-level effects impose two coupled requirements on bias-aware geometry design. First, the dependence of logical $Z$ and $X$ errors on lattice geometry and physical error rate must be characterized. Second, the logical response to a bias acting only on selected circuit locations must be distinguished from the response produced by an overall variation of the circuit error rate. In the CNOT-based syndrome-extraction circuits considered here, we adopt an HBD-based noise prescription in which the prescribed dephasing bias is retained only at explicit idle locations, while gates, measurements, and resets remain governed by unbiased, operation-dependent fault rates \cite{q7w6-nljp,gidney2021fault,gidney2022benchmarking}. Under this noise prescription, the nominal physical bias alone is insufficient to determine the resulting logical error asymmetry. Direct biased circuit-level simulations can reliably evaluate individual configurations, but independently simulating and decoding every geometry--bias condition becomes increasingly demanding as the design space expands. This motivates the use of a transferable prediction model for candidate screening, while direct circuit-level simulation is retained for final validation.

The central question of this work is whether the logical response to such a localized circuit-level bias can be inferred from an unbiased reference circuit and transferred to finite-bias conditions without geometry- or bias-specific response fitting. We consider unrotated rectangular surface codes under standard depolarizing (SD) and superconducting-inspired (SI) circuit-level noise models. Unbiased baseline models $F_Z(L_1,L_2,p)$ and $F_X(L_1,L_2,p)$ are first constructed to characterize the dependence of the logical error rates on lattice geometry and physical error rate. The effect of the idle-only bias is then incorporated through separate effective error parameters, $p_Z^{\mathrm{eff}}$ and $p_X^{\mathrm{eff}}$, for logical $Z$ and $X$ errors, respectively.

Recent logical-sensitivity approaches provide mechanism-resolved measures of the response of logical failure probabilities to variations in detector-error probabilities \cite{fu2026sensitivity,shao2026attribution}. These sensitivities, however, do not by themselves establish a mapping between a localized perturbation acting on a subset of circuit locations and an equivalent variation of the global physical error parameter used in a baseline logical error model. To establish such a mapping, we compare the logical responses along two physically distinct perturbation directions: a localized idle-bias perturbation and a global variation of the baseline physical error rate. Their relative response defines the logical-error-type-dependent coefficient $w_s^{(0)}$. The coefficients $w_Z^{(0)}$ and $w_X^{(0)}$ are evaluated entirely from an unbiased reference circuit, without fitting to finite-bias logical error rate data, and are subsequently used to construct the effective error parameters. The resulting predictions achieve a median fold error of $1.09$, compared with $1.28$ when the bias response is neglected and $3.18$ when the full idle-channel change is transferred to the effective baseline parameter.

Finally, the bias-dependent predictions are used to identify minimum-overhead rectangular geometries subject to a target logical error rate, and the selected geometries are independently validated by direct biased circuit-level simulations. At $p=0.004$ and $P_{L,\mathrm{target}}=10^{-3}$, the validated rectangular codes reduce physical qubit overhead by up to $42.4\%$ under the SD model and $26.2\%$ under the SI model relative to the corresponding unrotated square code baselines over the tested bias ratios $\eta\in\{1,2,3\}$. These results demonstrate, within the investigated regime, a practical connection between unbiased circuit characterization, localized bias-response estimation, and resource-efficient geometry selection under biased circuit-level noise.

\begin{figure}[b]

  \centering
  \includegraphics[width=1.0\textwidth]{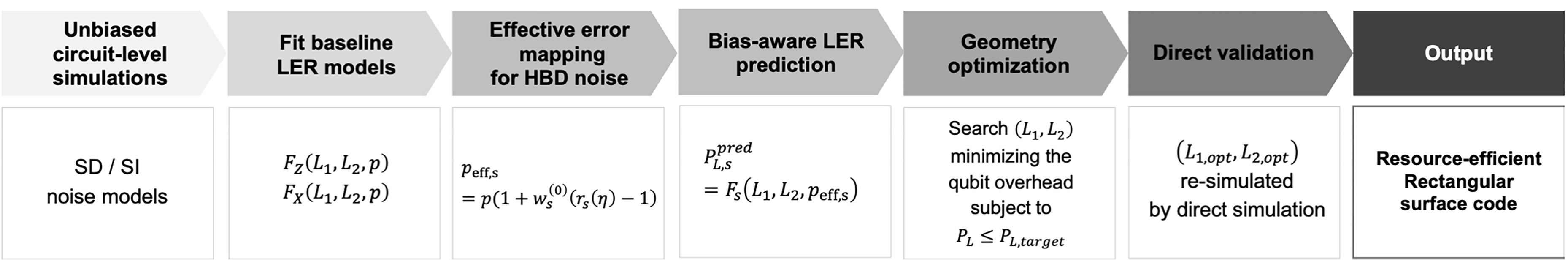}
  \caption{\textbf{Overall workflow of the rectangular surface code design framework.}
  Baseline logical error rate models are constructed from unbiased SD and SI circuit-level simulation data and extended to biased noise through an effective error mapping. The resulting bias-dependent model identifies the minimum-overhead lattice geometry among candidates predicted to satisfy the target logical error rate. The selected geometry is then independently evaluated using direct biased circuit-level simulations to verify target satisfaction.}
\label{fig:framework}
\end{figure}

\section{Results}\label{sec:results}

The overall workflow of the proposed rectangular surface code design framework is summarized in Fig.~\ref{fig:framework}. First, baseline logical error rate (LER) models are constructed from unbiased circuit-level simulations under the SD and SI noise models. An effective-error mapping is then constructed using response coefficients obtained from an unbiased reference circuit, without fitting to biased LER data, and applied to the baseline models for bias-aware LER prediction and geometry optimization. Finally, the selected lattice sizes are independently evaluated by direct biased circuit-level simulations to validate their logical performance and resource efficiency.

\begin{figure}[h]
\centering
\includegraphics[width=1.0\columnwidth]{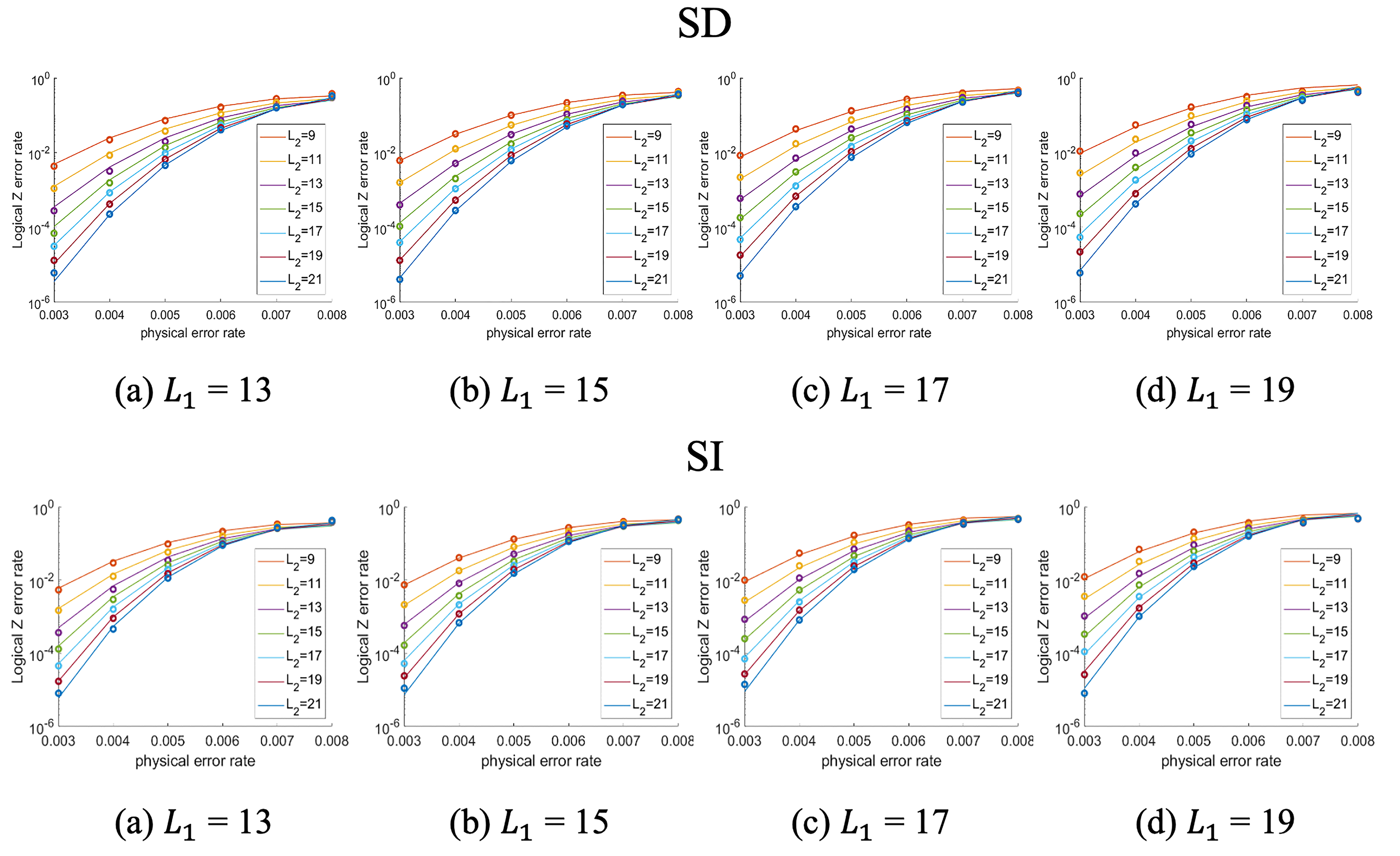}
\caption{
\textbf{Logical $Z$ error rates under the Standard Depolarizing
(SD) and Superconducting-Inspired (SI) circuit-level noise models.}
Results are shown for fixed $L_1$ values of
(a)~13, (b)~15, (c)~17 and (d)~19,
with $L_2 \in \{9, 11, 13, 15, 17, 19, 21\}$.
Markers represent the simulation data, and lines represent
the fitted logical $Z$ error rates obtained from
\eqref{eq:plz_log_model}.
A longer logical $Z$ operator ($L_2$) leads to improved
suppression of logical $Z$ errors in the low physical
error rate regime.
}
\label{fig:sd_plz}
\end{figure}

\subsection{Baseline logical error rate models}\label{subsec:ler_modeling}

Fig.~\ref{fig:sd_plz} shows the simulated logical $Z$ error rates and the corresponding fitted curves under the SD and SI circuit-level noise models.
To characterize the dependence of the logical error rate
on lattice geometry and the physical error rate, we
construct an empirical baseline model.
Motivated by the approximately exponential suppression
of logical errors with increasing code distance in the
low-error regime, we model $\ln P_{L_Z}$ using a term
linear in $L_2$ with a $p$-dependent coefficient.
Additional terms in $L_1$ and $\ln L_2$ are included
to capture residual geometry-dependent behavior.
The resulting expression is

\begin{equation}
    \begin{aligned}
    \ln P_{L_Z} & =
    \beta_0
    + \beta_1 \ln p
    + \beta_2(\ln p)^2  \\
    &+
    \left\{
    \beta_3
    + \beta_4 \ln p
    + \beta_5(\ln p)^2
    + \beta_6 p
    \right\}L_2
    + \beta_7 L_1
    + \beta_8 \ln L_2 .
    \end{aligned}
    \label{eq:plz_log_model}
\end{equation}

Here, $\beta_i$ are $p$-independent coefficients
determined by fitting the unbiased circuit-level
simulation data.
This expression serves as an empirical surrogate
for logical error rate prediction rather than a
theoretical scaling law, and the individual regression
coefficients are not assigned independent physical
interpretations.

The fitting accuracy of the baseline model is
evaluated against the simulated logical error rates.
For lattice geometries beyond the fitting domain,
the model is used to screen candidate configurations,
while the final resource comparisons are established
through independent direct circuit-level simulations.

Similarly, the logical $X$ error rate is obtained by modeling the scaling behavior along the code's horizontal direction, as the logical $X$ operator connects smooth boundaries. The resulting expression is

\begin{equation}
    \begin{aligned}
     \ln P_{L_X}
    =
    &\beta_0^{(x)}
    +\left\{
        \beta_1^{(x)}
        +\beta_3^{(x)}\ln p
        +\beta_4^{(x)}(\ln p)^2
        +\beta_5^{(x)}p
    \right\}L_1 \\
    &+
    \left\{
        \beta_2^{(x)}
        +\beta_6^{(x)}\ln p
        +\beta_7^{(x)}(\ln p)^2
    \right\}L_2
    +\beta_8^{(x)}\ln L_2 .
    \end{aligned}
    \label{eq:plx_model}
\end{equation}

Here, $\beta_i$ and $\beta_i^{(x)}$ denote $p$-independent regression coefficients for the logical $Z$ and $X$ error rate models, respectively. Both expressions serve as empirical surrogates for logical error rate prediction rather than theoretical scaling laws, and their individual regression coefficients are not assigned independent physical interpretations.

The coefficients were estimated by ordinary
least-squares regression using the unbiased
circuit-level simulation data.
Specifically, $\ln P_{L_Z}$ was used as the
response variable, and the corresponding design
matrix was constructed from $L_1$, $L_2$, $p$,
and $\ln p$ according to Eq.~\eqref{eq:plz_log_model}.
Because the model is linear in the coefficients
after the logarithmic transformation,
the coefficients were estimated by minimizing
the residual sum of squares between the
simulated and fitted values of $\ln P_{L_Z}$.

Similarly, the coefficients $\beta_i^{(x)}$
were obtained by applying the same least-squares
procedure to the simulated logical $X$
error rate data using Eq.~\eqref{eq:plx_model}.
The same fitting procedure was applied separately
to the SD and SI circuit-level noise models.
Further details of the logical $X$ model and
its fitting results are provided in the
Supplementary Note on logical $X$ error rate estimation.
The fitted coefficients for the SD and SI models
are listed in Supplementary Table S2.

Across the unbiased fitting domain,
the empirical baseline models reproduce
the simulated logical error rates with
$\log_{10}$-scale root-mean-square errors
below 0.06 for all four combinations of
noise model and logical error type.
The corresponding median fold errors
range from 1.048 to 1.087, indicating
typical multiplicative deviations of
approximately 5--9\% between the fitted
and simulated logical error rates.
These results show that, within the fitted
parameter range, the same empirical modeling
framework provides a consistent description
of both logical error types under the SD
and SI circuit-level noise prescriptions.
Detailed fitting statistics are provided
in Supplementary Table~S3.

Eqs.~\eqref{eq:plz_log_model} and~\eqref{eq:plx_model} therefore provide empirical baseline models that quantitatively describe the geometry-dependent logical error rates within the fitted domain under unbiased circuit-level noise. We further use the same models, without geometry-specific refitting, to extrapolate beyond the fitted lattice range and guide candidate selection. To avoid relying solely on these extrapolations, the final logical error rates and physical qubit comparisons reported in Section~\ref{subsec:geometry_validation} are established by independent direct circuit-level simulations. In the following section, these unbiased baseline models are extended to biased circuit-level noise through response coefficients obtained independently from an unbiased reference circuit.

\subsection{Prediction of biased logical error rates from an unbiased reference circuit}
\label{subsec:w0_prediction}

The baseline models in Eqs.~\eqref{eq:plz_log_model} and~\eqref{eq:plx_model}
are parameterized by the physical error rate parameter $p$ used in the unbiased
SD and SI circuit-level simulations.
In the hybrid biased--depolarizing (HBD)-based noise model (Section~\ref{subsec:noise_models}), however, the $\eta$-dependent Pauli asymmetry is
introduced only at explicit idle locations.
Therefore, directly replacing $p$ in the baseline models by one of the Pauli
components of the biased idle channel would incorrectly apply the idle-channel
asymmetry to all circuit fault locations.
Our goal is therefore to represent the effect of the idle-only physical bias as separate equivalent shifts of the baseline error rate parameter for logical Z and X errors, while determining this mapping from an unbiased reference circuit without calibrating its response coefficients to biased LER data.

We therefore introduce an effective-error mapping that converts the prescribed idle-noise bias into separate equivalent baseline error parameters for the logical $Z$ and $X$ error models.
For the idle-channel Pauli components in Eq.~\eqref{eq:idle_pauli_components}, the
relevant Pauli combinations for the logical $Z$ and $X$ errors are
$\pi_Z+\pi_Y$ and $\pi_X+\pi_Y$, respectively.
At the unbiased point $\eta=1/2$, both reduce to $2q_{\mathrm{idle}}/3$.
Their relative changes are therefore

\begin{equation}
r_Z(\eta)
=
\frac{\pi_Z+\pi_Y}{2q_{\mathrm{idle}}/3}
=
\frac{3(2\eta+1)}{4(\eta+1)},
\qquad
r_X(\eta)
=
\frac{\pi_X+\pi_Y}{2q_{\mathrm{idle}}/3}
=
\frac{3}{2(\eta+1)}.
\label{eq:r_zx}
\end{equation}

Here, $r_s(\eta)$ describes the relative change of the Pauli error
components relevant to logical error type $s\in{Z,X}$ at the physical idle channel level.
To translate this local change into the baseline error rate parameter that
characterizes the complete circuit, we define

\begin{equation}
p_s^{\mathrm{eff}}(p,\eta)
=
p
\left[
1+
w_s^{(0)}
\left(
r_s(\eta)-1
\right)
\right],
\qquad
s\in\{Z,X\}.
\label{eq:effective_p}
\end{equation}

The central quantities in Eq.~\eqref{eq:effective_p} are the response coefficients
$w_Z^{(0)}$ and $w_X^{(0)}$.
Unlike $r_s(\eta)$, which is determined by the Pauli composition of the
biased idle channel, $w_s^{(0)}$ accounts for how strongly this prescribed
idle-bias perturbation is reflected in the logical response of the complete
circuit.
Crucially, these coefficients are not obtained by fitting
Eq.~\eqref{eq:effective_p} to biased logical error rate data.
Instead, they are evaluated from an unbiased reference circuit at
$\eta=1/2$, $p=0.004$, and $L_1=L_2=13$.
For each logical error type $s$, we define

\begin{equation}
w_s^{(0)}
=
\frac{
\sum_i
\varepsilon_{i,s}b_{i,s}
}{
\sum_i
\varepsilon_{i,s}\kappa_{i,s}
},
\label{eq:w0}
\end{equation}

where $i$ indexes detector error mechanisms and $q_i$ denotes
the probability of mechanism $i$.
The logarithmic sensitivity
$\varepsilon_{i,s}=\partial\ln P_{L,s}/\partial\ln q_i$
quantifies the response of the logical failure probability to
a relative change in $q_i$ \cite{fu2026sensitivity,shao2026attribution}.
The quantities $b_{i,s}$ and $\kappa_{i,s}$ describe the responses
of $\ln q_i$ to the prescribed idle-bias perturbation
$\delta_s=r_s(\eta)-1$ and to $\ln p$, respectively.
All derivatives are evaluated at the same unbiased reference circuit,
with the decoder held fixed when defining $\varepsilon_{i,s}$.
The numerator therefore represents the local logical response
to the idle-bias perturbation, whereas the denominator represents
the response to a fractional change in the baseline physical
error rate.
The precise definitions and numerical estimation procedure are
given in Section~4.4 and the Supplementary Note on the estimation
of the reference-circuit response coefficients.

The resulting coefficients are $(w_Z^{(0)},w_X^{(0)}) = (0.3575,\,0.3831)$ for the SD model and $(w_Z^{(0)},w_X^{(0)})
= (0.2559,\,0.2386)$ for the SI model.
The corresponding statistical uncertainties are of order $10^{-3}$.
These coefficients are all substantially smaller than unity, showing that applying the
full relative change of the biased idle channel directly to the baseline error parameter
would overestimate its influence on the complete circuit.
However, $w_s^{(0)}$ is not interpreted as a literal fraction of physical bias
transmitted to the logical level.
It instead quantifies the local response of the logical error probability to the specified
idle-channel perturbation relative to an overall variation of the baseline error parameter.

Once $w_s^{(0)}$ is determined, the coefficients are kept fixed.
The biased logical error rates are predicted by evaluating the unbiased baseline models, Eqs.~\eqref{eq:plz_log_model} and~\eqref{eq:plx_model}, at the
corresponding effective error parameters :

\begin{equation}
\begin{aligned}
P_{L_Z}^{\mathrm{pred}}
&=
F_Z\!\left(L_1,L_2,p_Z^{\mathrm{eff}}\right),
P_{L_X}^{\mathrm{pred}}
=
F_X\!\left(L_1,L_2,p_X^{\mathrm{eff}}\right),
\\
P_L^{\mathrm{pred}}
&\approx
1-
\left(1-P_{L_Z}^{\mathrm{pred}}\right)
\left(1-P_{L_X}^{\mathrm{pred}}\right)
=
P_{L_Z}^{\mathrm{pred}}
+
P_{L_X}^{\mathrm{pred}}
-
P_{L_Z}^{\mathrm{pred}}P_{L_X}^{\mathrm{pred}} .
\end{aligned}
\label{eq:bias_ler}
\end{equation}

We evaluated this fixed-coefficient model against direct HBD circuit-level simulations at finite bias. To assess whether the response obtained from an unbiased reference circuit remains predictive when transferred to different lattice geometries and finite-bias conditions, we used a matched 132-point evaluation set. 
The evaluation set comprises 66 conditions per noise model, matched in $(\eta,L_1,L_2,p)$, and spans $\eta\in\{3,5\}$, nine lattice geometries, and $0.003\leq p\leq0.008$.

The $w_s^{(0)}$-based predictions give
a median fold error of $1.09$.
In contrast, neglecting the bias response by setting $w_s^{(0)}=0$ gives a median fold error
of $1.28$, whereas assuming that the full idle-channel change is transferred to the
effective baseline parameter, $w_s^{(0)}=1$, increases the median fold error to $3.18$.
These results indicate that the response obtained from an unbiased reference circuit can be transferred across the evaluated finite-bias and lattice-geometry conditions without geometry- or bias-specific recalibration, with substantially smaller prediction error than either neglecting the bias response or assuming its full transfer. 
Furthermore, excluding all conditions associated with the $13\times13$ reference geometry leaves 112 validation conditions, for which the fixed-coefficient model achieves an MFE of 1.077
(SD: 1.091; SI: 1.072). These results assess the predictive performance of the fixed-coefficient
model on geometries other than the reference configuration, with accuracy varying across geometries as shown in Supplementary Table~S6.
The prediction errors reflect the combined performance of the empirical baseline models and the response mapping; the present comparison
does not isolate their respective error contributions.

We therefore treat $w_s^{(0)}$ as a fixed reference parameter within the evaluated domain, not as a universal circuit constant; its transferability is assessed empirically through the logical error rate predictions over the matched finite-bias and lattice-geometry conditions above. Additional assessments of the fixed-coefficient predictions, including results resolved by bias, logical error type, and lattice geometry, are provided in Supplementary Tables~S4--S6.

In the following section we use the fixed mapping to identify candidate rectangular geometries predicted to satisfy the target logical error rate, and then independently verify their target satisfaction through direct biased circuit-level simulations.

\subsection{Geometry optimization and direct circuit-level validation}
\label{subsec:geometry_validation}

To determine the lattice size $(L_{1,\mathrm{opt}},L_{2,\mathrm{opt}})$
for a given target logical error rate
$P_{L,\mathrm{target}}$, physical error rate $p$, and bias $\eta$, we minimized the physical qubit overhead subject to the logical error rate predicted by the fixed-$w_s^{(0)}$ bias-dependent LER estimator in Eq.~\eqref{eq:bias_ler}. The optimization was performed by an exhaustive search over odd lattice
dimensions
$L_1,L_2\in\{7,9,\ldots,49\}$ :
\begin{equation}
\begin{aligned}
\underset{L_1,\,L_2}{\operatorname*{arg\,min}}\quad & 4L_1L_2 - 2L_1 - 2L_2 + 1 \\
\text{s.t.}\quad & P_L(L_1,L_2,p,\eta) \le P_{L,\mathrm{target}} .
\end{aligned}
\label{eq:geometry_optimization}
\end{equation}
Here, $P_L$ denotes the logical error rate estimated using Eq.~\eqref{eq:bias_ler}.

Accordingly, $(L_{1,\mathrm{opt}},L_{2,\mathrm{opt}})$
denotes the minimum-overhead lattice geometry within the specified search space that is predicted to satisfy the target logical error rate. Because the empirical LER model is used to rank candidate geometries, particularly when the selected sizes extend beyond the baseline fitting range, the model prediction alone is not used as the final criterion for target satisfaction.
The resulting candidate geometries were therefore subsequently evaluated using independent direct biased circuit-level simulations, as described in the Supplementary Note on direct simulation and geometry extrapolation validation.

\begin{table*}[t]
\caption{
\textbf{Direct-simulation comparison of selected rectangular surface codes and square code baselines.}
Results are shown for the SD and SI circuit-level noise models at $p\in\{0.003,0.004\}$ and $P_{L,\mathrm{target}}\in\{10^{-3},10^{-4}\}$. The rectangular geometries were selected using the
$w_s^{(0)}$-based logical error rate estimator and subsequently
evaluated through direct biased circuit-level simulations.
The square codes serve as reference baselines under the
corresponding noise conditions.
The reported logical error rates are obtained from direct
memory experiments with $r=\max(L_1,L_2)$ syndrome-extraction rounds.
The corresponding 95\% confidence intervals are reported in Supplementary Table~S7.
}
\label{tab:optimized_validation}

\centering
\small
\renewcommand{\arraystretch}{1.15}
\setlength{\tabcolsep}{2pt}


\begin{tabular*}{\textwidth}{
@{\extracolsep{\fill}}
l c c
c c c c
c c c
}
\toprule
\multicolumn{10}{c}{\textbf{(a) SD circuit-level noise model}} \\
\midrule

\multicolumn{3}{c}{} &
\multicolumn{4}{c}{\textbf{Rectangular code}} &
\multicolumn{3}{c}{\textbf{Square code}} \\
\cmidrule(lr){4-7}
\cmidrule(l){8-10}

\textbf{$P_{L,\mathrm{target}}$}
& \textbf{$\eta$}
& \textbf{$p$}
& $\boldsymbol{L_1\!\times\!L_2}$
& \shortstack{\textbf{\# of}\\\textbf{qubits}}
& \textbf{$P_L^{\mathrm{pred}}$}
& \textbf{$P_L^{\mathrm{direct}}$}
& $\boldsymbol{L\!\times\!L}$
& \shortstack{\textbf{\# of}\\\textbf{qubits}}
& \textbf{$P_L^{\mathrm{direct}}$} \\
\midrule

\multirow{4}{*}{$10^{-3}$}
& 1 & 0.004
& $21\times23$ & 1845
& $8.98\times10^{-4}$
& $8.58\times10^{-4}$
& $23\times23$ & 2025
& $7.38\times10^{-4}$ \\

& 2 & 0.004
& $19\times25$ & 1813
& $7.47\times10^{-4}$
& $7.14\times10^{-4}$
& $25\times25$ & 2401
& $7.34\times10^{-4}$ \\

& 3 & 0.004
& $17\times25$ & 1617
& $8.81\times10^{-4}$
& $8.23\times10^{-4}$
& $27\times27$ & 2809
& $7.00\times10^{-4}$ \\

& 3 & 0.003
& $11\times13$ & 525
& $9.09\times10^{-4}$
& $7.75\times10^{-4}$
& $13\times13$ & 625
& $7.68\times10^{-4}$ \\

\multirow{1}{*}{$10^{-4}$}
& 3 & 0.003

& $15\times19$ & 1073
& $7.69\times10^{-5}$
& $7.46\times10^{-5}$
& $19\times19$ & 1369
& $6.40\times10^{-5}$ \\

\bottomrule
\end{tabular*}

\vspace{1.2em}


\begin{tabular*}{\textwidth}{
@{\extracolsep{\fill}}
l c c
c c c c
c c c
}
\toprule
\multicolumn{10}{c}{\textbf{(b) SI circuit-level noise model}} \\
\midrule

\multicolumn{3}{c}{} &
\multicolumn{4}{c}{\textbf{Rectangular code}} &
\multicolumn{3}{c}{\textbf{Square code}} \\
\cmidrule(lr){4-7}
\cmidrule(l){8-10}

\textbf{$P_{L,\mathrm{target}}$}
& \textbf{$\eta$}
& \textbf{$p$}
& $\boldsymbol{L_1\!\times\!L_2}$
& \shortstack{\textbf{\# of}\\\textbf{qubits}}
& \textbf{$P_L^{\mathrm{pred}}$}
& \textbf{$P_L^{\mathrm{direct}}$}
& $\boldsymbol{L\!\times\!L}$
& \shortstack{\textbf{\# of}\\\textbf{qubits}}
& \textbf{$P_L^{\mathrm{direct}}$} \\
\midrule

\multirow{4}{*}{$10^{-3}$}
& 1 & 0.004
& $25\times29$ & 2793
& $8.93\times10^{-4}$
& $9.3\times10^{-4}$
& $27\times27$ & 2809
& $9.30\times10^{-4}$ \\

& 2 & 0.004
& $23\times29$ & 2565
& $9.63\times10^{-4}$
& $9.73\times10^{-4}$
& $29\times29$ & 3249
& $6.50\times10^{-4}$ \\

& 3 & 0.004
& $23\times31$ & 2745
& $7.88\times10^{-4}$
& $7.85\times10^{-4}$
& $31\times31$ & 3721
& $8.53\times10^{-4}$ \\

& 3 & 0.003
& $13\times15$ & 725
& $5.99\times10^{-4}$
& $5.17\times10^{-4}$
& $15\times15$ & 841
& $4.80\times10^{-4}$ \\

\multirow{1}{*}{$10^{-4}$}
& 3 & 0.003
& $17\times21$ & 1353
& $6.77\times10^{-5}$
& $5.10\times10^{-5}$
& $21\times21$ & 1681
& $4.80\times10^{-5}$ \\

\bottomrule
\end{tabular*}

\end{table*}

We then compared the physical qubit counts of the model-selected rectangular codes and the square baselines satisfying a fixed target logical error rate, as summarized in Table~\ref{tab:optimized_validation}.
We first considered $P_{L,\mathrm{target}}=10^{-3}$ at
$p=0.004$ and bias values of $\eta=1,2,$ and $3$.
The rectangular geometries reported in
Table~\ref{tab:optimized_validation} are the minimum-qubit solutions among the candidate geometries predicted by Eq.~(\ref{eq:geometry_optimization}) to satisfy the target logical error rate. Their target satisfaction was subsequently confirmed by independent direct circuit-level simulations using the upper bounds of the 95\% confidence intervals.
The square surface codes were evaluated under the same SD or SI noise model as
the corresponding rectangular codes, and each square baseline is the lowest-overhead directly validated square code satisfying the same condition.
Additional lower-overhead geometries were directly simulated to assess prediction accuracy, with the results reported in the Supplementary Note on direct simulation and geometry extrapolation validation.

\begin{figure}[t]
\centering
{\includegraphics[width=0.7\columnwidth]{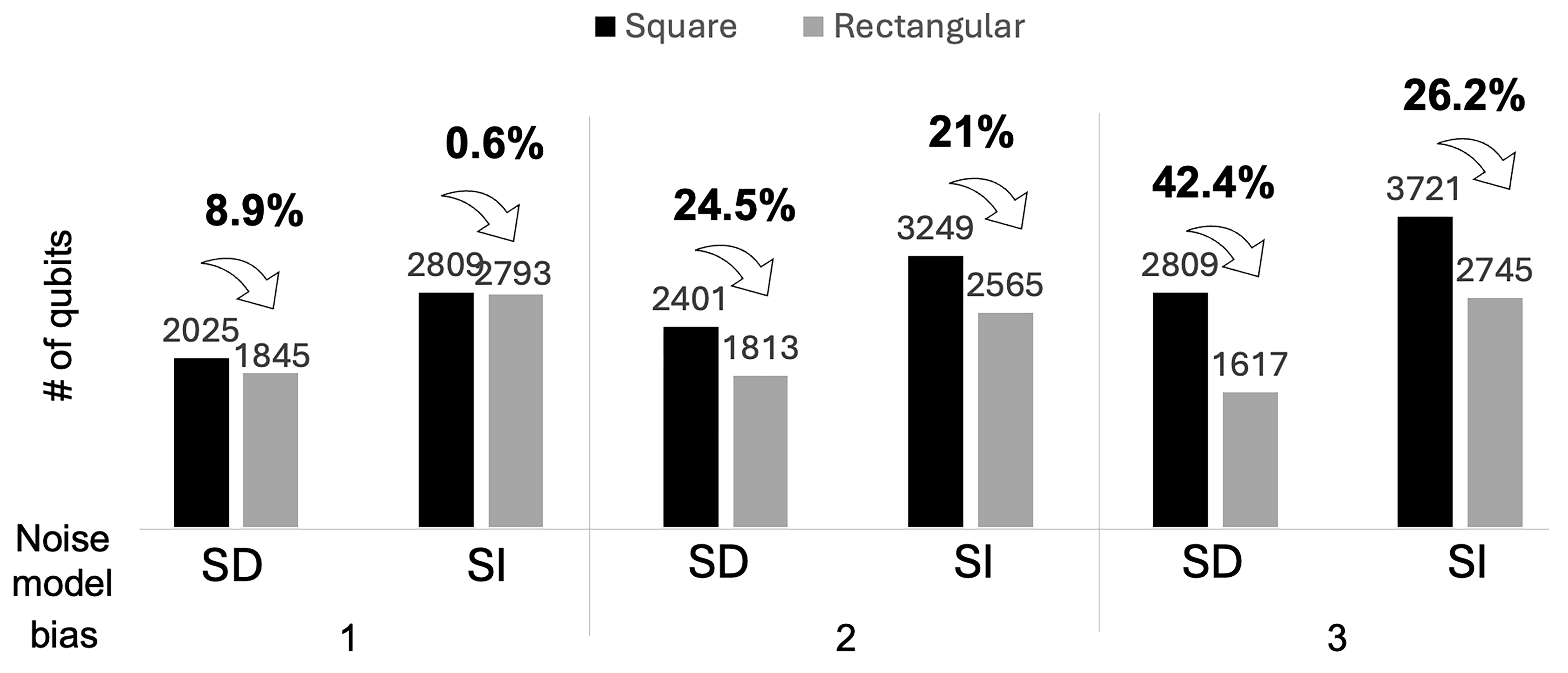}}
\caption{
\textbf{Physical qubit reduction achieved by rectangular surface codes.}
Comparison of the physical qubit overhead of the rectangular geometries and corresponding square baselines listed in Table~\ref{tab:optimized_validation} at $p=0.004$ and $P_{L,\mathrm{target}}=10^{-3}$.
Rectangular codes require fewer qubits than the corresponding square baselines, with reductions of up to approximately 42.4\% under the SD model and
26.2\% under the SI model over the bias values considered.
}
\label{fig:qubit_reduction}
\end{figure}

The implementation of rectangular surface codes leads to a substantial reduction
in the total physical qubit overhead under the biased circuit-level conditions
considered here (Fig.~\ref{fig:qubit_reduction}).
For the SD model, the physical qubit reduction relative to the corresponding
square baseline increases from approximately 8.9\% at $\eta=1$ to 24.5\% at
$\eta=2$ and 42.4\% at $\eta=3$.
At $\eta=3$, the $17\times25$ rectangular code uses 1617 physical qubits and
yields $P_L=8.23\times10^{-4}$, whereas the corresponding $27\times27$
square code requires 2809 qubits and yields
$P_L=7.00\times10^{-4}$.

Under the SI model, the reduction is smaller at weak bias but becomes more
pronounced as the bias increases.
The rectangular codes reduce the physical qubit count by approximately 0.6\%,
21\%, and 26.2\% at $\eta=1,2,$ and $3$, respectively.
At $\eta=3$, the $23\times31$ rectangular geometry uses 2745 physical qubits
and yields $P_L=7.85\times10^{-4}$.
In contrast, the corresponding $31\times31$ square baseline requires 3721
physical qubits and yields $P_L=8.53\times10^{-4}$.

The SI model generally requires larger lattices and higher total qubit counts
than the SD model at the same $p$, $\eta$, and target logical error rate.
For example, at $\eta=2$ and $p=0.004$, the selected SI rectangular code
requires 2565 physical qubits, whereas the corresponding SD rectangular code
requires 1813 qubits.
This difference reflects the combined effect of the operation-dependent fault
rates in the SI prescription, including its different reset, measurement, gate,
and idle error probabilities, rather than a uniform rescaling of the SD model.
These results emphasize that operation-dependent circuit-level fault assumptions
can materially affect the resource-efficient lattice geometry.

The resource advantage is not limited to the primary operating point of
$p=0.004$ and $P_{L,\mathrm{target}}=10^{-3}$.
At $p=0.003$ and $\eta=3$, the directly validated rectangular geometries
continue to require fewer physical qubits than the corresponding square
baselines for both $P_{L,\mathrm{target}}=10^{-3}$ and $10^{-4}$.
These additional operating points also show that the selected lattice geometry
depends on both the physical error rate and the target logical error rate.
The corresponding 95\% confidence intervals are reported in Supplementary Table~S7.

\begin{table*}[h]
\caption{\textbf{Logical performance at comparable physical qubit overhead.}
Directly simulated logical error rates are compared between rectangular and square surface codes under the SD and SI noise models.
For each case, the square lattice uses a comparable or slightly larger number of physical qubits than the corresponding rectangular lattice.}
\label{tab:comparable_overhead}

\centering
\small
\renewcommand{\arraystretch}{1.15}
\setlength{\tabcolsep}{1.5pt}

\begin{tabular*}{\textwidth}{
@{\extracolsep{\fill}}
c c c c c c c c c c
}
\toprule

$\eta$
& $p$
& \makecell{Rectangular\\$L_1 \times L_2$}
& \makecell{\#\\qubits}
& \makecell{$P_{L,\mathrm{rect}}$\\SD}
& \makecell{$P_{L,\mathrm{rect}}$\\SI}
& \makecell{Square\\$L \times L$}
& \makecell{\#\\qubits}
& \makecell{$P_{L,\mathrm{sq}}$\\SD}
& \makecell{$P_{L,\mathrm{sq}}$\\SI} \\
\midrule

\multirow{2}{*}{3}
& 0.005
& $13\times17$
& 825
& $3.28\times10^{-2}$
& $6.67\times10^{-2}$
& $15\times15$
& 841
& $4.26\times10^{-2}$
& $7.21\times10^{-2}$ \\

& 0.004
& $11\times15$
& 609
& $6.22\times10^{-3}$
& $1.26\times10^{-2}$
& $13\times13$
& 625
& $8.41\times10^{-3}$
& $1.28\times10^{-2}$ \\

\midrule

\multirow{2}{*}{2}
& 0.005
& $15\times19$
& 1073
& $2.84\times10^{-2}$
& $6.55\times10^{-2}$
& $17\times17$
& 1089
& $3.48\times10^{-2}$
& $6.71\times10^{-2}$ \\

& 0.004
& $17\times21$
& 1353
& $1.51\times10^{-3}$
& $3.75\times10^{-3}$
& $19\times19$
& 1369
& $2.28\times10^{-3}$
& $4.59\times10^{-3}$ \\

\bottomrule
\end{tabular*}

\end{table*}

As a complementary analysis, Table~\ref{tab:comparable_overhead} compares rectangular and square surface codes with comparable physical qubit overhead.
In this comparison, the square lattices were chosen to use a similar or slightly
larger number of physical qubits than the corresponding rectangular lattices.
For the conditions examined in Table~\ref{tab:comparable_overhead}, the rectangular codes show lower directly simulated logical error rates than the corresponding square
baselines under both the SD and SI noise models.
For example, at $\eta=2$ and $p=0.004$, the $17\times21$ rectangular code uses 1353 qubits and yields
$P_L=1.51\times10^{-3}$ under the SD model and
$P_L=3.75\times10^{-3}$ under the SI model.
In comparison, the $19\times19$ square code uses 1369 qubits but yields higher logical error rates of
$P_L=2.28\times10^{-3}$ and $P_L=4.59\times10^{-3}$ under the SD and SI models, respectively.
These results provide a complementary indication that, at comparable qubit cost, rectangular geometries can improve logical performance under the investigated
hybrid biased circuit-level noise conditions.

\section{Discussion}\label{sec:discussion_conclusion}

In this work, we investigated resource-efficient rectangular surface code design under biased circuit-level noise, focusing on whether the logical response to a localized physical bias can be inferred from an unbiased reference circuit. The unbiased baseline models described the geometry-dependent logical $Z$ and $X$ error rates under the SD and SI noise models with median fold errors of approximately 1.05--1.09. We then introduced separate effective error parameters for logical $Z$ and $X$ errors using the response coefficients $w_Z^{(0)}$ and $w_X^{(0)}$, evaluated from an unbiased reference circuit rather than fitted to finite-bias data. Despite being determined solely from the unbiased reference circuit, the fixed-coefficient model achieved a median fold error of 1.09 over the evaluated finite-bias conditions. For the same finite-bias evaluation set, the zero-response assumption, $w_s^{(0)}=0$, and the full-response assumption, $w_s^{(0)}=1$, yielded median fold errors of 1.28 and 3.18, respectively. These results indicate that neither neglecting the bias response nor assuming its full transfer adequately captures the logical response of the complete circuit.

This comparison also clarifies why the nominal physical bias alone is insufficient to determine logical performance under circuit-level noise. In the HBD-based prescription used here, the prescribed Pauli asymmetry is applied only to explicit idle locations, whereas gate, reset, and measurement faults retain their corresponding unbiased circuit-level channels \cite{q7w6-nljp,meinking2026leveraging}. The resulting coefficients, $(w_Z^{(0)},w_X^{(0)})=(0.3575,0.3831)$ for SD and $(0.2559,0.2386)$ for SI, are all substantially below unity. From Eq.~\eqref{eq:w0}, $w_s^{(0)}$ should not be interpreted as a literal fraction of the physical bias transmitted to the logical level. Instead, it quantifies the sensitivity-weighted logical response of the bias-responsive idle mechanisms relative to the response produced by an overall variation of the baseline physical error parameter. This construction uses mechanism-resolved logical sensitivities to establish a directional response mapping that connects an unbiased reference circuit to finite-bias logical error prediction. The systematically smaller coefficients obtained for SI are therefore consistent with a smaller relative contribution of the biased idle subset under its operation-dependent fault prescription.

Applying this fixed mapping to geometry selection resulted in substantial physical qubit savings relative to square codes. At $p=0.004$ and $P_{L,\mathrm{target}}=10^{-3}$, the reduction increased with bias and reached 42.4\% for SD and 26.2\% for SI at $\eta=3$. Smaller reductions of 13.8–21.6\% were obtained at the additional $p = 0.003$ operating points under the same acceptance criterion. The selected geometries were independently verified by direct circuit-level simulations, with target satisfaction assessed using the upper bounds of the corresponding 95\% confidence intervals. Thus, the reported resource reductions do not rely solely on extrapolation of the empirical LER model.

Taken together, these results suggest a practical separation between two components of biased-noise surface code design: the geometry dependence of the logical error rate in an unbiased circuit, described by $F_Z(L_1,L_2,p)$ and $F_X(L_1,L_2,p)$, and the response of that circuit to a localized bias perturbation, summarized by $w_Z^{(0)}$ and $w_X^{(0)}$. Within the range studied here, combining these quantities enabled finite-bias LER prediction and geometry selection without separately fitting the bias response for every $\eta$ and lattice geometry. This separation is particularly relevant at the circuit level, where physical bias may be retained only at a subset of fault locations rather than uniformly across the syndrome-extraction circuit \cite{q7w6-nljp,gidney2021fault,gidney2022benchmarking}.

Several limitations define the scope of these conclusions. The response coefficients were obtained from a single $13\times13$ reference circuit at $p=0.004$, and their transferability was evaluated only over the noise parameters and lattice geometries considered here. They should therefore be regarded as fixed reference parameters within the tested regime rather than universal constants. A new reference characterization may be required when the syndrome-extraction circuit or operation-dependent noise prescription changes, or when the set of circuit locations carrying the physical bias differs materially from that used to determine $w_s^{(0)}$. In addition, the present simulations employ stochastic Pauli fault channels and do not include leakage, coherent errors, or temporally correlated noise that may arise in realistic quantum devices \cite{harper2023learning,mcewen2021removing}. The study is also restricted to unrotated rectangular surface codes with MWPM decoding.

Future work should examine the transferability of the reference-circuit response across different code distances, circuit constructions, and distributions of biased locations. Extending the framework to rotated surface codes \cite{xiao2024exact} and more realistic hardware noise models would further test its generality. More broadly, the present results indicate that circuit-level noise characterization can be connected to resource-efficient code design through a compact intermediate response model, while direct simulation remains the final validation step for the selected geometry.

\section{Methods}\label{sec:methods}
\subsection{Circuit-level noise models}\label{subsec:noise_models}

We consider two circuit-level noise prescriptions, the standard depolarizing (SD) and superconducting-inspired (SI) models \cite{gidney2021fault,gidney2022benchmarking}. Both models assign stochastic faults to the elementary operations of the syndrome-extraction circuit, including gates, reset, measurement, and idling, but differ in how the corresponding fault rates are distributed across operation types. In the SD model, all operations are assigned the same reference fault rate $p$. In contrast, the SI model assigns operation-dependent fault rates to account for the different error tendencies associated with the durations of the corresponding circuit operations. The fault rates used in this work are summarized in Table~\ref{tab:noise_models}.

In quantum systems, Pauli $X$, $Y$, and $Z$ errors do not necessarily occur with equal probabilities, and one type of error may dominate the others \cite{iolius2022performance,seis2023improving}. In particular, in dephasing-dominated environments, phase-flip ($Z$) errors occur more frequently than $X$ or $Y$ errors. Such asymmetry is referred to as biased noise. We describe biased noise using an asymmetric Pauli channel with error probabilities $p_X, p_Y$, and $p_Z$, and  define the bias ratio as $\eta=p_Z/(p_X+p_Y)$. A depolarizing channel
($p_X=p_Y=p_Z$) corresponds to $\eta=1/2$, whereas larger values of $\eta$ correspond to increasingly dephasing-dominated noise.

For biased circuit-level simulations, we employ an HBD-based noise model \cite{q7w6-nljp}. In two-level qubit systems, gate operations such as Hadamard and CNOT do not generally preserve the underlying dephasing bias throughout syndrome extraction. Applying the same biased Pauli channel to all circuit locations can therefore overestimate the amount of bias retained at the circuit level. Following the HBD framework, we retain the dephasing bias only at explicit idle locations.

\begin{table}[h]
\caption{\textbf{Circuit-level fault rates for the SD and SI noise models.}}
\label{tab:noise_models}

\centering
\renewcommand{\arraystretch}{1.05}

\begin{tabular}{@{}lcc@{}}
\toprule
\textbf{Qubit operation}
& \makecell{\textbf{Standard}\\\textbf{Depolarizing (SD)}}
& \makecell{\textbf{Superconducting-}\\\textbf{Inspired (SI)}} \\
\midrule
CNOT & $p$ & $p$ \\
Hadamard ($H$) & $p$ & $p/10$ \\
Reset to $\ket{0}$ ($\ket{+}$) & $p$ & $2p$ \\
Measurement & $p$ & $5p$ \\
Idle during gates on other qubits & $p$ & $p/10$ \\
Idle during reset or measurement of other qubits & $p$ & $2p$ \\
\botrule
\end{tabular}

\end{table}

Idle faults were separately assigned to qubits idling during gate layers and to qubits idling during measurement/reset periods, with the corresponding total fault probabilities determined by the SD or SI rates in Table~\ref{tab:noise_models}.
Specifically, the bias parameter $\eta$ modifies only the relative Pauli composition of an idle fault, while its total fault probability remains determined by the corresponding SD or SI prescription.
Denoting the total fault probability of an idle channel by $q_{\mathrm{idle}}$, the Pauli components are assigned as
\begin{equation}
\pi_X=\pi_Y=\frac{q_{\mathrm{idle}}}{2(\eta+1)},
\qquad
\pi_Z=\frac{\eta q_{\mathrm{idle}}}{\eta+1}.
\label{eq:idle_pauli_components}
\end{equation}
The unbiased case corresponds to $\eta=1/2$, for which $\pi_X=\pi_Y=\pi_Z=q_{\mathrm{idle}}/3$. This unbiased setting is used for both baseline LER model fitting and the estimation of the reference circuit response coefficients $w_s^{(0)}$.
Hadamard and CNOT faults are modeled by symmetric one- and two-qubit depolarizing channels, respectively, while reset and measurement faults retain their SD- or SI-specific rates without additional bias reweighting.
Further details of the syndrome-extraction schedule and circuit-level fault assignments are provided in the Supplementary Note on circuit-level implementation of the biased noise models.

\begin{figure}[h]
\centering
\includegraphics[width=0.65\columnwidth]{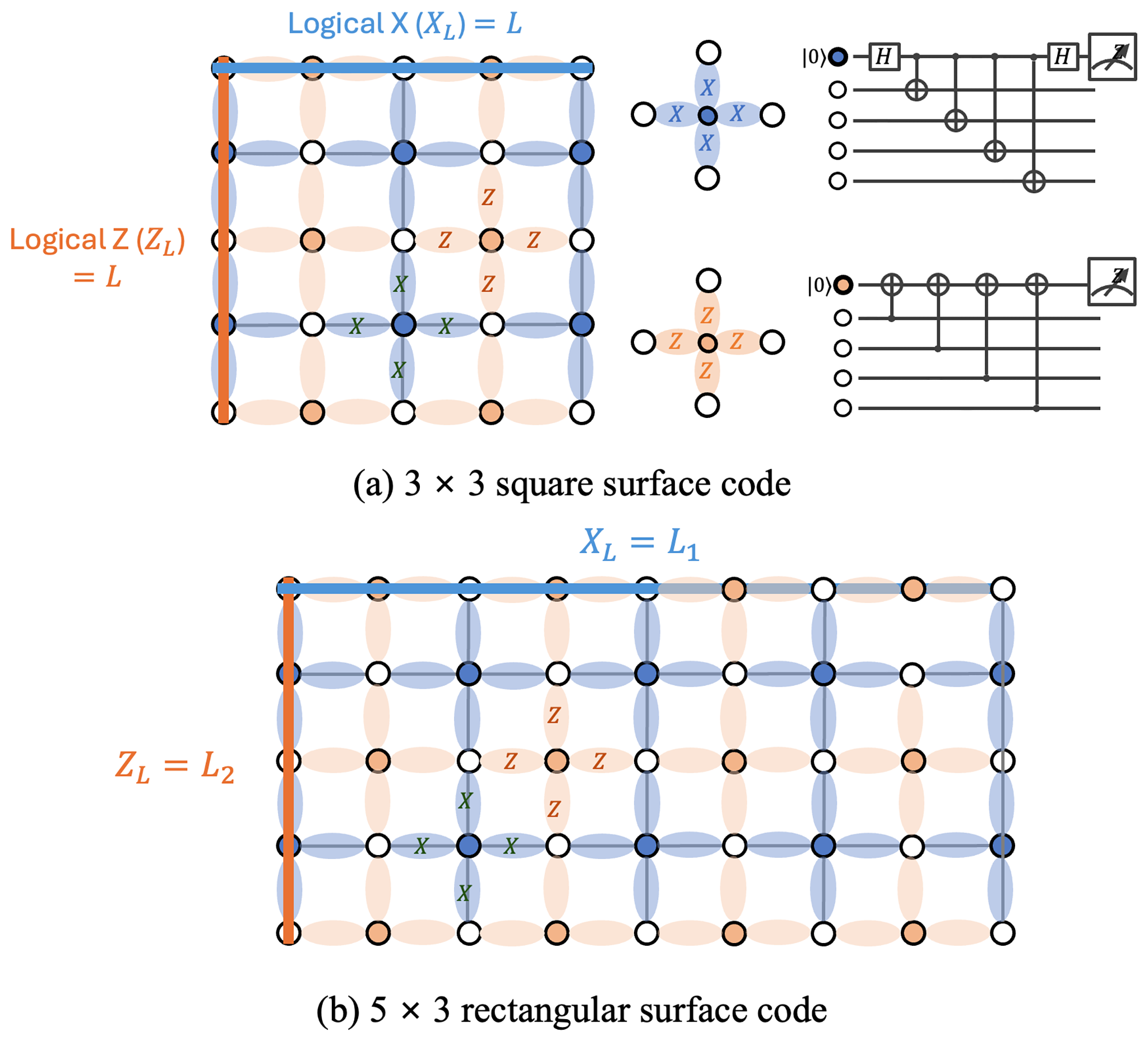}
\caption{
\textbf{Square and rectangular surface code geometries and stabilizer circuits.}
Illustration of (a) logical operators and stabilizer measurement circuits in  $3\times3$ square and (b) $5\times3$ rectangular surface codes. On the left, the logical operators $X_L$ and $Z_L$ are shown as strings of Pauli operators 
that connect the boundaries of the same type. The blue ($X$-type) vertex and orange ($Z$-type) plaquettes correspond to stabilizers that detect phase-flip and bit-flip errors, respectively. On the right, the corresponding quantum 
circuits depict how $X$ and $Z$ stabilizers are measured.
}
\label{fig:surface_code}
\end{figure}

\subsection{Rectangular surface code}\label{subsec:rectangular_code}

We consider an unrotated planar surface code defined on an asymmetric $L_1 \times L_2$ lattice (Fig. \ref{fig:surface_code}). The minimum weights of the logical $X$ and logical $Z$ operators are $L_1$ and $L_2$, respectively, allowing the two logical distances to be adjusted independently. A square surface code corresponds to $L_1=L_2$, whereas a rectangular geometry enables different levels of protection against the two types of logical errors \cite{lee2021rectangular}.

Under dephasing-dominated noise, choosing $L_2>L_1$ increases the logical $Z$ distance and provides stronger protection against logical $Z$ failures while avoiding unnecessary protection along the less affected logical direction. This anisotropic degree of freedom motivates the use of rectangular surface codes under biased noise \cite{lee2021rectangular}. For an $L_1 \times L_2$ unrotated rectangular surface code, the total number of physical qubits, including both data and measurement qubits, is $N_{\mathrm{phys}}=4L_1L_2-2L_1-2L_2+1$. For a square surface code with $L_1=L_2=L$, this reduces to $N_{\mathrm{phys}}=(2L-1)^2$. 

\subsection{Numerical simulations}
\label{subsec:simulation_methods}

In our analysis, the logical error rate is characterized
in terms of the lattice sizes ($L_1$ and $L_2$) and
the physical error rate $p$.
We simulated unrotated rectangular surface code
memory experiments under the SD and SI
circuit-level noise models.
Simulations were performed using Stim
\cite{gidney2021stim}, a tool for modeling and
analysis of stabilizer circuits;
PyMatching \cite{higgott2022pymatching},
a fast minimum-weight perfect-matching (MWPM)
decoder; and Sinter \cite{gidney2021stim} for multiprocessing-based
batch sampling and decoding of Stim circuits.

To construct the circuits, we adapted the built-in
generators of Stim such that whenever specific
qubits were idle during a given time step,
explicit idling faults were applied to those qubits.
Each generated circuit was converted into a
detector error model, and the resulting syndrome
data were decoded using PyMatching on the
corresponding detector graph.

To construct the baseline LER models, unbiased
simulations ($\eta=1/2$) were performed over
the physical error rate range
$0.003 \leq p \leq 0.008$,
with $L_1 \in \{13,15,17,19\}$ and
$L_2 \in \{9,11,13,15,17,19,21\}$.
The baseline model coefficients were determined
by ordinary least-squares regression of the
logarithmic logical error rates using the
corresponding model basis functions.
Independent fits were performed for the logical
$Z$ and $X$ errors under each noise model.
Only unbiased simulation data were used for
baseline LER fitting, without calibration
using finite-bias LER data.

To reduce the statistical uncertainty of the
estimated logical error rates, Monte Carlo
sampling was performed using Sinter.
For each simulation point, at least $10^6$
shots were collected.
For all memory experiments, including the
unbiased simulations used to construct the
baseline LER models and the direct biased
simulations, the number of syndrome extraction
rounds was set to $r=\max(L_1,L_2)$,
which reduces to $r=L$ for square $L\times L$
codes.
Therefore, all logical error rates reported
in this work correspond to the full
memory-experiment failure probability over
$r$ syndrome extraction rounds.
The direct biased circuit-level simulations
used the same circuit construction, sampling,
and decoding procedure, with the HBD-based
noise model described in
Section~\ref{subsec:noise_models}.

\subsection{Estimation of reference-circuit response coefficients}
\label{subsec:reference_response}

The logical-error-type-dependent response coefficients $w_s^{(0)}$,
with $s\in\{Z,X\}$, were evaluated from an unbiased
reference circuit for each noise model. Specifically,
we employed an unrotated $13\times13$ surface code
memory circuit at $p=0.004$ and $\eta=1/2$.
The reference circuit was converted into a detector
error model (DEM), where each error mechanism $i$
has an associated probability $q_i$.

We first considered the relative changes in the idle-channel Pauli components relevant to logical $Z$ and $X$ errors,
defined as $\delta_s=r_s(\eta)-1$, where $r_s(\eta)$
is given in Eq.~(\ref{eq:r_zx}). This perturbation changes only
the Pauli composition of explicit idle faults while
keeping their total fault probabilities fixed.
The corresponding logarithmic response of each
DEM mechanism was defined as
\begin{equation}
{b}_{i,s}
=
\left.
\frac{\partial \ln q_i}
{\partial \delta_s}
\right|_{\delta_s=0}.
\end{equation}
The derivatives were estimated using central finite
differences of the DEM mechanism probabilities at
$\eta=0.5\pm0.005$.

To quantify the effect of these mechanism-level
changes on the logical error rate, we evaluated
the logarithmic logical sensitivity
${\varepsilon}_{i,s}
=\partial\ln P_{L,s}/\partial\ln q_i$.
The sensitivities were estimated from Monte Carlo
samples of the unbiased DEM using a fixed MWPM
decoder constructed from the reference circuit.
Let $N_{F,s}$ denote the number of logical failures
and $H_{i,s}$ the number of those failures in which
mechanism $i$ occurred. The corresponding
score-function estimator is
\begin{equation}
\widehat{{\varepsilon}}_{i,s}
=
\frac{H_{i,s}/N_{F,s}-q_i}{1-q_i}.
\label{eq:10}
\end{equation}
This procedure estimates the sensitivities of
individual DEM mechanisms without separately
perturbing each mechanism. Following \cite{fu2026sensitivity,shao2026attribution}, Eq.~\ref{eq:10} provides mechanism-resolved logical sensitivities from a common set of Monte Carlo samples. In the present work, these sensitivities are subsequently combined with the responses of the detector-error mechanisms to two physically distinct perturbations, corresponding to a localized variation of the idle-channel bias and a fractional variation of the baseline physical error rate. This construction allows the first-order logical responses along the two perturbation directions to be compared within the same reference circuit.

The second perturbation direction is defined by a fractional variation of the baseline physical error rate, $p\rightarrow p(1+\delta_p)$, which changes the operation-dependent fault probabilities according to the SD or SI prescription. The corresponding
logarithmic response is defined as
\begin{equation}
{\kappa}_{i,s}
=
\left.
\frac{\partial\ln q_i}
{\partial\ln p}
\right|_{p=0.004},
\end{equation}
and is estimated using central finite differences
with $p\rightarrow p(1\pm0.005)$.
Using the same logical sensitivities, the first-order logical response along each perturbation direction is obtained by weighting the corresponding mechanism-level responses by ${\varepsilon}_{i,s}$.
The response coefficient $w_s^{(0)}$ is then defined by the ratio of these two logical responses, as given in Eq.~(\ref{eq:w0}). Accordingly, $w_s^{(0)}$ quantifies the relative logical response to the localized idle-bias perturbation with respect to a fractional variation of the baseline physical error rate and maps the former onto an equivalent variation of the baseline error parameter.

For each noise model and logical error type,
$5\times10^6$ shots were sampled and divided
into 40 independent chunks. The statistical
standard error of $w_s^{(0)}$ was estimated using
a leave-one-chunk-out jackknife procedure.
Once determined, the coefficients were kept fixed
for finite-bias LER prediction and geometry
optimization, without fitting to biased
logical error rate data. Further numerical
details of the sensitivity estimation
and finite-difference response calculations are
provided in the Supplementary
Information.

\subsection{Statistical analysis and validation}
\label{subsec:statistical_analysis}

Logical error rates were estimated from Monte Carlo
simulations of logical $Z$ and $X$ memory experiments.
For each logical error type $s\in\{Z,X\}$, the logical error rate
was estimated as $\widehat{P}_{Ls}=k_s/N_s$,
where $k_s$ denotes the number of observed logical
failures and $N_s$ is the total number of shots.
The combined logical error rate was approximated as
$\widehat{P}_L=1-(1-\widehat{P}_{LZ})
(1-\widehat{P}_{LX})$.
Statistical uncertainties were quantified using the
Clopper--Pearson exact confidence intervals.
A Bonferroni correction was applied to obtain
a simultaneous confidence level of at least 95\%,
corresponding to two-sided 97.5\% confidence intervals
for each logical error type.
The confidence limits of the combined logical error rate were obtained by applying the same combination formula to the corresponding lower and upper confidence bounds for logical $Z$ and $X$ errors.

Prediction accuracy was evaluated using the median
fold error (MFE), defined as
\begin{equation}
\mathrm{MFE}
=
\operatorname{median}_i
\left[
10^{\left|
\log_{10}P_{L,i}^{\mathrm{pred}}
-
\log_{10}P_{L,i}^{\mathrm{direct}}
\right|}
\right],
\label{eq:mfe}
\end{equation}
where $P_{L,i}^{\mathrm{pred}}$ and
$P_{L,i}^{\mathrm{direct}}$ denote the predicted
and directly simulated logical error rates,
respectively.
Baseline fitting accuracy was assessed using
unbiased simulation data, whereas finite-bias
prediction accuracy was evaluated over 132 conditions,
comprising 66 conditions for each noise model.
The same evaluation set was used to compare
the predictions obtained with $w_s^{(0)}=0$, $w_s^{(0)}=1$,
and $w_s^{(0)}$.
The reference coefficients $w_s^{(0)}$ were determined
exclusively from unbiased reference circuits
and remained fixed throughout the finite-bias
prediction evaluations.

For geometry optimization, candidate lattice sizes
were evaluated using the predicted logical error
rates and physical qubit overhead.
Selected rectangular geometries were subsequently
validated through independent biased circuit-level
simulations.
A geometry was considered to satisfy the target
logical error rate only when the upper bound of
its combined 95\% confidence interval did not exceed
$P_{L,\mathrm{target}}$.
The square code baselines were determined using
the same statistical criterion.
Resource efficiency was compared between the model-selected rectangular geometries and the square baselines, both of which satisfied this criterion.

\bmhead{AI-assisted manuscript preparation}
ChatGPT (OpenAI) was used to assist with English-language editing, manuscript formatting, and preliminary review of statistical descriptions.
The authors retain full responsibility for all scientific analyses, interpretations, and conclusions.

\bmhead{Code availability}
The custom code used for circuit generation,
simulation, and response-coefficient estimation
is available from the corresponding author
upon reasonable request.

\bmhead{Data Availability}
The datasets generated and/or analysed during
the current study are available from the
corresponding author upon reasonable request.

\backmatter

\bibliography{references}

\bmhead{Acknowledgements}
This work was supported by the Institute of Information \& Communications Technology Planning \& Evaluation (IITP), Information Technology Research Center (ITRC) grant funded by the Korean government (Ministry of Science and ICT) (IITP-2026-RS-2021-II211810).

\bmhead{Author contributions}
Y. Chung and J. Heo developed the theoretical idea. Y. Chung performed the main analysis and wrote the manuscript. Y. Kang provided technical support. All authors contributed to the analysis of the results.

\bmhead{Competing interests}
The authors declare no competing interests.

\bmhead{Additional information}
Supplementary Information accompanies this manuscript as a separate document.

\end{document}